\documentclass[aps,prd,twocolumn]{revtex4-2}
\usepackage[english]{babel} 
\usepackage{amssymb}
\usepackage{amsmath}
\usepackage{txfonts}
\usepackage{mathdots}
\usepackage{graphicx}
\usepackage{subfig}
\usepackage{bm}

\newcommand{\Y}{\Y_{lm}}

\newcommand{\be}{\begin{equation}}
\newcommand{\ee}{\end{equation}}
\newcommand{\Be}{\begin{eqnarray}}
\newcommand{\Ee}{\end{eqnarray}}

\newcommand{\f}{\frac}

\begin{document}
\pagestyle{plain}

\title{Wormhole Effective Mass and Gravitational Waves
by Binary Systems Containing Wormhole}

\author{Sung-Won Kim}
\email[email:]{sungwon@ewha.ac.kr}
\affiliation{Ewha Womans University, Seoul 03760, Korea}

\begin{abstract}
We considered the generation of gravitational waves by the binary system associated with a wormhole. In the Newtonian limit, the gravitational potential of a wormhole requires the effective mass of the wormhole taking into account radial tension effects. This definition allows us to derive gravitational wave production in homogeneous and heterogeneous binary systems. Therefore, we studied gravitational waves generation by orbiting wormhole-wormhole and wormhole-black hole binary systems before coalescence. Cases involving negative mass require more careful handling.
We also calculated the energy loss to gravitational radiation by a particle orbiting around the wormhole and by a particle moving straight through the wormhole mouth, respectively.
\end{abstract}

\maketitle
\date{today}
\section{Introduction}

LIGO/VIRGO's success  \cite{Aasi,Ace} in detecting gravitational waves has ushered in a new era. There are many reports on gravitational wave sources and templates. Typically, compact objects such as black holes and/or neutron stars are the primary sources, but exotic objects such as wormholes can also be astrophysical compact objects that are candidates for gravitational wave sources\cite{TBBL2021}, even though their existence is still not clear.

However, we believe that there is enough value to consider gravitational waves caused by wormholes or wormhole-black hole pairs. There are numerous attempts to trace wormholes, such as gravitational lensing \cite{Abe2011,TKA2011}, shadows \cite{NTY2013}, Einstein rings \cite{THY2012} and particle creation \cite{Kim1992}. If the detection of gravitational waves generated by a wormhole or any system associating with wormhole is successful, this detection may also be added to the list of wormhole evidence.

As a first step in gravitational waves generated by a wormhole, we recently studied a toy model of a rotating thin-shell wormhole \cite{kim2023}. The amplitude of gravitational waves caused by perturbed precession  of a slowly rotating, thin-shell wormhole is very small. Moreover, the mass and energy of a thin-shelled wormhole are negative.
This means that negative energy radiation reduces the size of the wormhole. However, the lifetime of the wormhole is longer than cosmological  time.

Binary system is currently the most likely source of gravitational waves.
Like conventional gravitational wave sources, wormholes constitute binary systems and are sufficient to generate gravitational waves in the system. It also allows us to deal with wormhole-black hole binaries before merging, even without knowing in detail the final state of the heterogeneous binary system of similar mass as the merging mechanism.

The final states in the extreme cases can be inferred relatively easily. For example, if the black hole mass is extremely small compared to the wormhole mass, which will be discussed later, the black hole is a particle that orbits the wormhole along a geodesic line and ultimately passes through the wormhole while causing a small disturbance to the wormhole.
When the black hole mass is extremely larger than the wormhole mass, a wormhole is a particle that orbits a black hole and is ultimately absorbed into the black hole like the other particle.

There have been several studies on binary systems including wormholes. Very early on, a simulation of two wormholes merging was attempted \cite{HL1964}. Their original problem was two masses interacting in the framework of geometrodynamics. They identified the particles as multi-connected regions of empty space.
Cardoso et al. \cite{CFP2016} considered the final state of two wormholes merging and a wormhole binary ringdown state through quasi normal modes (QNM) analysis. Even without an event horizon, the QNM exists with light rings.
There was a study on gravitational waves by a black hole orbiting a wormhole \cite{DGHK21}.
They considered a very small mass of black hole comparing to the thin-shell wormhole mass.
They found a characteristic waveform-an antichirp and/or burst-as the black hole put spirals into our region of the Universe.
A recent study was the wormhole-black hole collision by Dias et al \cite{DFPR2023}. They used ray-tracing techniques to analyze the event horizon of a large black hole with a very small wormhole.

In this paper, we study the inspiral stage of a binary system before the coalescence stage. We considered a homogeneous binary system (e.g. wormhole-wormhole) and a heterogeneous binary system (e.g. wormhole-blackhole). After proper mass analysis, we studied the waveforms and energy loss due to gravitational wave generation in wormhole binary systems.
Finally, as an extreme case of a binary system, we got gravitational waves generated by small particles around the wormhole and energy loss from particles traveling directly into the wormhole mouth.

In Section II, we review the definition and physical implications of wormhole mass. Next, in Section III, we treat the gravitational potential of a wormhole, deriving the effective mass of the wormhole from Newtonian approximation. In Section IV, we study the generation of gravitational waves by binary systems containing wormholes, such as wormhole-wormhole, wormhole-black hole, and a point particle-wormhole.

\section{Mass of a Wormhole}

To check the gravitational interaction around a wormhole, let's trace the case of an ordinary star. If two massive objects are far enough apart, they can both be considered point particles. When nearby, gravitational interactions depend on the distribution of matter and the associated structures.
There may also be finite-size effects, such as tidal effects caused by nearby gravitational fields.

The most convenient model of the wormhole is the Morris-Thorne type as \cite{MT}
\be
ds^2=-e^{2\Phi(r)}c^2dt^2+ \left(1 - \frac{b(r)}{r}\right)^{-1}dr^2+r^2(d\theta^2+\sin^2 \theta d\phi^2),
\ee
where $\Phi(r)$ is the red-shift function and $b(r)$ is the wormhole shape function. These two functions are determined by the wormhole-consisting materials through the Einstein's equation as
\Be
b' &=& 8\pi G c^{-2}r^2\rho, \label{eq:b-rho}\\
\Phi'&=& \frac{b-8\pi Gc^{-4}\tau r^3}{2r\left(r-b\right)}, \label{eq:pot}\\
\tau'&=& \left(\rho c^2-\tau \right)\Phi'-{2\left(P+\tau \right)}/r.
\Ee
Here $\rho$ is the mass density, $\tau$ is the surface tension which is the negative radial pressure, and $P$ is the transverse pressure. The main condition imposed on this wormhole matter is the flare-out condition as
\be
\zeta \equiv \frac{\tau -\rho c^2}{\left|\rho c^2\right|}>0 \label{eq:flare}
\ee
at or near throat, to main the shape of the wormhole.  This exotic matters violates the weak energy condition.

The most important problem with the wormhole is the existence of the negative density \cite{MT}. This is due to the flare-out condition (\ref{eq:flare}). The Lorentz transformation of this relation shows that the measurements by an observer passing through the throat at a radial velocity close to $c$, i.e. $\gamma \gg 1$, indicate a negative density.
\Be
T_{\hat{0'}\hat{0'}}&=& \gamma^2T_{\hat{t}\hat{t}}\mp 2\gamma^2 ({v}/{c})^2 T_{\hat{t}\hat{r}} + \gamma^2({v}/{c})^2 T_{\hat{r}\hat{r}}, \nonumber \\
&=&
\gamma^2 (\rho_0c^2-\tau_0)+\tau_0.
\Ee
The static observer may see negative density. So if we have no choice but to use the negative density, we are interested in minimizing the use of these exotic material.

The next question is whether wormholes can have negative mass, which is related to negative density. However, the density of a wormhole may be negative, but its mass may be positive. From (\ref{eq:b-rho}),
\be
b(r) = b(r_0) + \int^r_{r_0}{8\pi c^{-2}\rho(r') r'^2dr'} = 2 \f{Gm_w(r)}{c^2},\label{eq:b_mass}
\ee
where $m_w(r)$ is defined by \cite{Visser_book}
\be
m_w(r)\equiv \left(\frac{c^2b(r_0)}{2G}\right) + \int^r_{r_0}{4\pi \rho r'^2dr'}
\ee
as the wormhole mass inside the radius $r$. Here the density is distributed from $r_0$ to arbitrary $r$. The shape function $b(r)$ has a meaning of the mass distribution inside the wormhole \cite{Visser_book}. Thus in the case of spatial infinity distribution,
the mass
\[
\lim_{r\to\infty} m_w(r) = M_w
\]
is defined by a constant.

There are several examples where wormholes have negative density and negative mass.
The Ellis-Bronnikov wormhole \cite{Ellis,Bron} shows negative density when the first derivative of the shape function is negative due to  asymptotic flatness in the power-law distribution.
We can also see the negative density of the wormhole model in lower dimensional spacetimes. The other example of the negative density is the case of the thin-shell wormhole. For this thin-shell wormhole,
 we cut out two copies of black holes outside the event horizon and paste them together like surgical \cite{Visser}.
The surface stress-energy tensor for the thin-shell is given by
\be
S_{ij}=\left( \begin{array}{ccc}
\sigma  & 0 & 0 \\
0 & p_{\vartheta } & 0 \\
0 & 0 & p_{\varphi } \end{array}
\right),
\ee
where $\sigma$ is the surface energy density, and $p_{\vartheta}$, $p_{\varphi}$ are principal surface pressures. The Einstein equation of the thin-shell wormhole is
\Be
\sigma &=& -\frac{1}{4\pi G}\left(\frac{1}{R_1}+\frac{1}{R_2}\right),\\
\vartheta_1 &=& -\frac{1}{4\pi G}\frac{1}{R_2},\\
\vartheta_2 &=& -\frac{1}{4\pi G}\frac{1}{R_1}.
\Ee
In general, since the junction region $\partial \Omega$ is convex,
namely, the principal curvatures $R_1, R_2$ are positive,
it has the negative surface energy density and negative surface tension.

\section{Gravitational potential and the effective mass of a wormhole}

To see the motion of particles around a wormhole, we first need to know the potential form of the wormhole.
The only constraints on potential $\Phi$ are that there is no horizon and that $\Phi(r)$ is everywhere finite \cite{MT}. If there is no cutoff in matter distribution, it is asymptotically flat, ${b}/{r}\to 0$ and $\Phi \to 0$  as  $r\to \infty$. To find the physical meaning of the potential, we need to analyze it using the Newtonian approximation.
Usually $\Phi$ is interpreted as the Newtonian gravitational potential $\Phi_N$
as
\be
\Phi \simeq -\f{\Phi_N}{c^2}
\ee
in weak field approximation.

Let us now look at the Newtonian approximation to the relativistic stellar potential. From the Einstein's equation for the relativistic star,
\be
\Phi'=\frac{4\pi Grp_r/c^4+Gm/c^2r^2}{\left(1-2Gm/rc^2\right)},
\ee
we get the potential for Newtonian star as
\be
 \Phi'_N=-\frac{Gm}{r^2},
\ee
by neglecting the pressure term in the numerator and the second term in the denominator. The neglected terms are smaller than the other terms by a factor of $1/c^2$.

Likewise, among the Einstein's equation for the wormhole, the potential-related equation (\ref{eq:pot})  is rewritten as
\be
\Phi'
=\frac{b/2r^2-4\pi G \tau r/c^4}{1- b/r}.
\ee
Here
we cannot neglect the pressure (tension) term in the numerator, unlike the relativistic star. This is because the pressure term cannot be neglected and larger than the density term due to the flare-out condition (\ref{eq:flare}). The second term in the denominator is neglected due to (\ref{eq:b_mass}).
Therefore the Newtonian approximation of the wormhole potential is
\be
 \Phi'_N=-\left(\frac{bc^2}{2r^2}-4\pi Gr\frac{\tau }{c^2}\right)=-\frac{Gm_{\mathrm{eff}}}{r^2}.
\ee
Here
\be
m_{\mathrm{eff}} \equiv \frac{bc^2}{2G}-\frac{4\pi}{c^2}\tau r^3
\ee
is the effective mass, which plays an important role in the gravitational potential $\Phi$. When considering gravitational interaction with a wormhole, the effective mass should be used as the wormhole mass in Newtonian limit. Therefore, this effective mass definition is applicable to any binary systems containing wormhole(s).

In the definition of effective mass, tension effect must be included in the mass of the normal materials. The sign of the effective mass
is determined by the magnitude of the tension.
The effective mass can be expressed in terms of wormhole materials, $\rho$ and $\tau$, as
\be
m_{\mathrm{eff}}=\frac{4\pi }{c^2}\left[\int{\rho c^2r^2dr}-\tau r^3\right].
\ee
The huge tension is required to keep the wormhole shape, by the flare-out condition, $\rho c^2 < \tau $.  Even if the tension is very large, the first term's range of integration can prevent the effective mass from taking on negative values. When $\tau>0$, the effective mass is smaller than the wormhole mass, $m_w=bc^2/2G$. And the effective mass for negative $\tau$ is larger than wormhole mass.
If the effective mass is independent of $r$, the wormhole can be considered as a point mass.
There are finite-size effects when the effective mass depends on $r$, i.e., when it is not uniform.

\section{gravitational waves by binary system}

\subsection{General formulas of gravitational waves by two-body system}

Before studying the wormhole binary system, we begin with the two-body motion of Kepler problem according to Newtonian analysis. They orbit each other under the influence of gravitation, but due to loss of gravitational wave energy, they get closer and eventually merge. If a black hole or neutron star is used as the body model, the final state will be the black hole state due to the high gravitational interaction.

The three stages in which a binary system settles into a stable state through orbital motion and coalescence are
inspiral, merger, and ringdown. Typically the first stage is treated analytically by the post-Newtonian approximation, and the merger stage is understand by the numerical analysis. The final ringdown stage is analyzed by quasi-normal modes.

We now restrict our study to the orbital-inspiral phase, dealing
with the Newtonian approximation ignoring the angular momentum term following the traditional derivations from the book \cite{CA_book}. The total energy of the binary system is
\be
E=\frac{1}{2}m_1v^2_1+\frac{1}{2}m_2v^2_2-\frac{Gm_1m_2}{a}=-\frac{1}{2}\mu v^2
\ee
for circular orbit. Here $\mu$ is the reduced mass and $a$ is the distance between the two masses.
The mass-related quantities are
\[
M=m_1+m_2, \quad \mu =\frac{m_1m_2}{M}, \quad \eta =\frac{\mu }{M}=\frac{m_1m_2}{M^2},
\]
where $M$ is the total mass, and $\eta$ is the symmetric mass ratio, equal to zero if one of the two masses is the test mass and 1/4 if both masses are the same mass.

From the Kepler's law
\be
GM=a^3\omega^2,
\ee
the tangent velocity is
\be
v=a\omega =\sqrt{\frac{GM}{a}}=\sqrt[3]{\frac{2\pi GM}{P}},
\ee
where $\omega$ is the angular frequency and $P$ is the period of the orbital motion.

The gravitational wave luminosity is
\be
L_{GW}=\f{32}{5}\f{c^5}{G}\eta^2 \left( \f{v}{c} \right)^{10}
=\f{32}{5}\f{c^5}{G}\left( \f{2G\mathcal{M}\omega}{c^3} \right)^{10/3},
\ee
where $\mathcal{M}=\eta^{3/5}M$ is the chirped mass.
Since $L_{GW}=-dE/dt$, we found that
\be
\f{dv}{dt}=\f{32\eta}{5}\f{v^9}{GMc^5}. \label{eq:velocity}
\ee
The time until coalescence, starting from orbital velocity $v_0$, is calculated 
as
\be
t_c=\frac{5}{256\eta }\frac{GM}{c^3}{\left(\frac{v_0}{c}\right)}^{-8}.
\ee
Here we assume that the two bodies merge at a distance $a \rightarrow 0$ including the wormhole case, and thus $v \rightarrow \infty$ at the coalescence time. However,  for a test particle with a wormhole, the closest  distance is the non-zero wormhole throat size.
Thus any time, in term of $v$, is
\be
t\left(v\right)
=t_c-\frac{5}{256\eta }\frac{GM}{c^3}{\left(\frac{v}{c}\right)}^{-8}.
\label{eq:time-vel}
\ee
The corresponding phase is derived 
as
\be
\varphi(t)
= \varphi_c - \f{c^5}{32\eta}v^{-5}=
{\varphi }_c-{\left[\frac{(t_c-t)c^3}{5G\mathcal{M}}\right]}^{\frac{5}{8}}.
\label{eq:phase}
\ee
Here $\varphi_c$ is the phase at coalescence and
the temporal change of the frequency is
\be
\frac{df}{dt}=\frac{96}{5}{\pi }^{8/3}{\left(\frac{c^3}{G\mathcal{M}}\right)}^{5/3}f^{11/3},
\ee
where $f=v^3/\pi GM$ is the gravitational frequency.
The frequency change rate for the wormhole can be compared to that for other objects, since the rate depends on mass.
Therefore the wave forms are
\begin{widetext}
\Be
h_+\left(t\right)&=& -\frac{G\mathcal{M}}{c^2r}\frac{1+{{\mathrm{cos}}^2 \iota \ }}{2}{\left(\frac{c^3(t_c-t)}{5G\mathcal{M}}\right)}^{-1/4}{\mathrm{cos} \left[2{\varphi }_c-2{\left(\frac{c^3(t_c-t)}{5G\mathcal{M}}\right)}^{5/8}\right]\ }, \\
h_{\times }\left(t\right) &=& -\frac{G\mathcal{M}}{c^2r}{\mathrm{cos} \iota \ }{\left(\frac{c^3(t_c-t)}{5G\mathcal{M}}\right)}^{-1/4}{\mathrm{cos} \left[2{\varphi }_c-2{\left(\frac{c^3(t_c-t)}{5G\mathcal{M}}\right)}^{5/8}\right]\ },
\Ee
\end{widetext}
where $\iota$ is the observation angle.
So we see that most of the quantities associated with gravitational waves, such as luminosity, wave forms, frequency change, and phase, can be represented by the chirped mass $\mathcal{M}$.

When applying angular momentum to Kepler problem, the effective potential must be defined as
\be
V_{\rm eff}=V_c+V_g=\frac{L^2}{2\mu r^2}-\frac{G\mu M}{r}.
\ee
Here $V_c$ is centrifugal potential with angular momentum $L$ and $V_g$ is the gravitational potential.
Unlike the traditional Kepler problem, the motion of a binary system
containing a wormhole is determined by the sign of its masses.

\begin{center}
\begin{figure}[b]
{{\includegraphics[width=1.63in, height=1.81in, keepaspectratio=false]{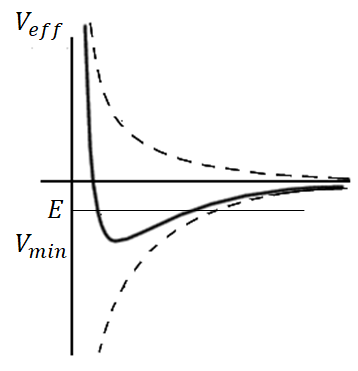}}}
{{\includegraphics[width=1.63in, height=1.81in, keepaspectratio=false]{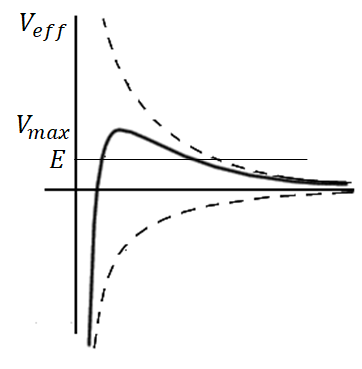}}}
\caption{Effective potentials (solid line) with the bounded motion. The dashed lines are $V_c$ or $V_g$. The left panel corresponds to case (1), and bounded motion is possible if the energy is $0>E>V_{\mathrm{min}}$. The right panel corresponds to the case (2), which allows bounded motion if the energy is $0<E<V_{\mathrm{max}}$.   }
\end{figure}
\end{center}

Because the mass can be taken as a negative or positive quantity,
these issues can be analyzed in the four cases: $m_1\lessgtr 0, m_2 \lessgtr 0$.
Of these, only two cases allow the bounded motion of the particle.
(1) If $m_1>0,~m_2>0$, then $\mu >0$ and the kinetic energy is positive. The centrifugal potential and the gravitational potential  are $V_c>0$, and $V_g<0$, respectively. Therefore, for negative energy larger than minimum value of the effective potential
($V_{\rm min}=-G^2M^2\mu^3/2L^2$), the particle has the bounded motion, while the particle with positive energy has unbounded motion.

(2) If $m_1>0,~m_2<0$, and $\mu <0$, then $V_c<0,~V_g>0$, and the effective potential has the inverted shape of case (1). Since the kinetic energy is negative, the particle motion is possible only in the region where the energy less than the effective potential energy. The motion is bounded when the total energy is between zero and the maximum value of the effective potential ($V_{\rm max}=-G^2M^2\mu^3/2L^2$).
As you can see, bounded motion is possible only for positive $M$ cases. For these, the effective potentials for cases (1) and (2) are shown in Fig. 1.

\subsection{Application to wormhole involved system}

From now on, we limit the mass of the wormhole to a constant effective mass  independent of $r$ to keep the problem simple.
If $\tau \propto r^{-3}$, the effective mass is a constant independent of $r$ and therefore can be considered as a point mass.
Let the density $\rho(r)$ be also proportional to $r^{-3}$ as
\be
\rho = \rho _0{\left(\frac{r_0}{r}\right)}^3,
\ee
where $\rho_0$ is the density at throat and $r_0$ is the lower limit of the matter distribution. The equation of state is assumed as
\be
\tau = k\rho c^2 = k \rho_0 c^2 {\left(\frac{r_0}{r}\right)}^3,
\ee
where $k$ is the dimensionless equation-of-state parameter and $k>1$ due to the flare-out condition (\ref{eq:flare}).
Then the shape function is given as
\be
b=\frac{8\pi G}{c^2}\int_{r_0}^r\rho(r') r'^2dr = \frac{8\pi G}{c^2}r^3_0
\rho_0 \ln \left( \f{r}{r_0} \right).
\ee
When the matter is infinitely distributed, $b$ diverges. To treat a wormhole as a point particle of constant mass, we must restrict
the distribution of matter to a special location $r_1$. The resulting effective mass is
\Be
m_{\mathrm{eff}} &=& \frac{bc^2}{2G}-4\pi r^3\frac{\tau }{c^2} = 4\pi \rho_0 r^3_0 \left[\ln \left(\frac{r_1}{r_0}\right) - k\right], \nonumber \\
&=& 4\pi \rho_0 r^3_0 (\ell -k). \label{eq:effective_mass}
\Ee
 Here $\ell=\ln(r_1/r_0)$ is a parameter of the mass density distribution range, and for the effective mass to be positive,  $\ell$ must be larger than $k$.
Set $m \equiv 4\pi\rho_0 r_0^3 \ell$ and $\alpha \equiv k/\ell$, where $\alpha$ is the ratio of tension to wormhole size and also decides the characteristics of the wormhole.

Let's use the model with constant effective mass and apply it to the practical problems of binary system including wormhole(s).
Examples here are wormhole-wormhole system, wormhole-black hole system, and particle-wormhole system.

\subsubsection{Wormhole-wormhole binary}

Consider the wormhole binary with effective masses and  a circular orbit using the Newtonian approximation. Substituting  $m_{\mathrm{eff}}$ for $m$ gives mass-related quantities $M_{\mathrm{eff}}$, $\mu_{\mathrm{eff}}$, $\eta_{\mathrm{eff}}$, and  $\mathcal{M}_{\mathrm{eff}}$, are defined respectively.
For example, the positive effective mass whose $\ell$ is sufficiently large gives the Hamiltonian and energy loss rate as
\be
H = -\frac{1}{2}\mu_{\mathrm{eff}} v^2,\quad L_{GW}=\frac{32c^2}{5G}\left(\frac{2G\mathcal{M}_{\mathrm{eff}}
\omega}{c^3}\right)^{10/3}.
\ee

The contribution of effective mass to the generation of gravitational waves depends on the mass included in the physical quantities related to gravitational waves. Therefore, it is necessary to check how dependent each physical quantity is on mass and to know how much it changes if they are replaced by effective mass.
For gravitational wave energy loss rate has the factor of
${\mathcal{M}}_{\mathrm{eff}}^{10/3}=\left[ (m_1m_2)^{3/5}/(m_1+m_2)^{1/5} \right]^{10/3}=(m_1m_2)^2/(m_1+m_2)^{2/3}$.

As can be seen from the cases mentioned above, the effective mass
can be either positive or negative. Therefore, the bounded orbital motion is possible only if the total effective mass is positive.
In other words, it is possible when both effective masses are positive,
or when one of them is negative but the total effective mass is positive.

Assume that a binary system  consists of two identical wormholes
with effective mass (\ref{eq:effective_mass}).
The mass-related quantities are
\[
m_\mathrm{eff} = m(1-\alpha),\quad M_\mathrm{eff} = 2m(1-\alpha)
 = 2m_\mathrm{eff}
,
\]
\[
\mu_\mathrm{eff}=\frac{m(1-\alpha)}{2}
=\frac{m_\mathrm{eff}}{2}
,
 \quad \eta_\mathrm{eff} =\frac{1}{4}.
\]
The chirped mass
\[{\mathcal{M}}_\mathrm{eff}=
(m_\mathrm{eff})^{6/5}
(M_\mathrm{eff})^{-1/5}=2^{-1/5}m(1-\alpha)
=2^{-1/5}m_{\mathrm{eff}}
.
\]
Here we want to deal with a positive effective mass wormhole binary, so we set $\alpha<1$.
We can now determine the effective mass effect of gravitational waves using the mass dependence formula.
When $\alpha=1/2$, the effective mass is the half the mass.
Let us assume that the initial conditions are the same as in the ineffective case.
The coalescence time is half that. In the formula (\ref{eq:time-vel}),
$v \propto M^{1/2}$, so $t_c-t \propto M^{-3}$ and $\varphi_c - \varphi \propto M^{-5/2}$.
Due to these mass dependencies, the waveform for effective mass is as shown in Fig. 2, and the case of ineffective mass such as a black hole binary or a neutron star binary is also drawn. To see the effect of effective mass, the amplitude was normalized excluding the mass term, and the chirp mass was set to 20, and the coalescence time was set to 100.
The ratio of the luminosity of twin wormhole binary with positive effective masses to the luminosity of normal star binary is
\be
\f{L_{GW}^{(\alpha)}}{L_{GW}}= (1-\alpha)^{10/3}.
\ee

As the time approaches the coalescence time, the velocity and phase grow faster than the ineffective mass. Because the coalescence time is halved, and the mass and time dependence is
$v \propto (M_{\mathrm{eff}}/(t_c-t))^{1/8}$, the rate of increase in velocity is larger.
As shown in (\ref{eq:phase}), $\varphi_c - \varphi \propto [(t_c-t)/m_{\mathrm{eff}}]^{5/8}$ and the rate of increase in the phase is also larger.

The left panel of Fig. 3 shows the ratio of wave amplitude and luminosity of a twin wormhole binary to those of twin black hole binary as a function of $\alpha$. 
As $\alpha$ increases, the amplitude and luminosity decrease.
$\alpha=1$ means that the effective mass of the wormhole is zero, there are no waves, and
there is no energy loss.

\begin{figure}
\includegraphics*[width=8.29cm, height=3.79cm, keepaspectratio=false]{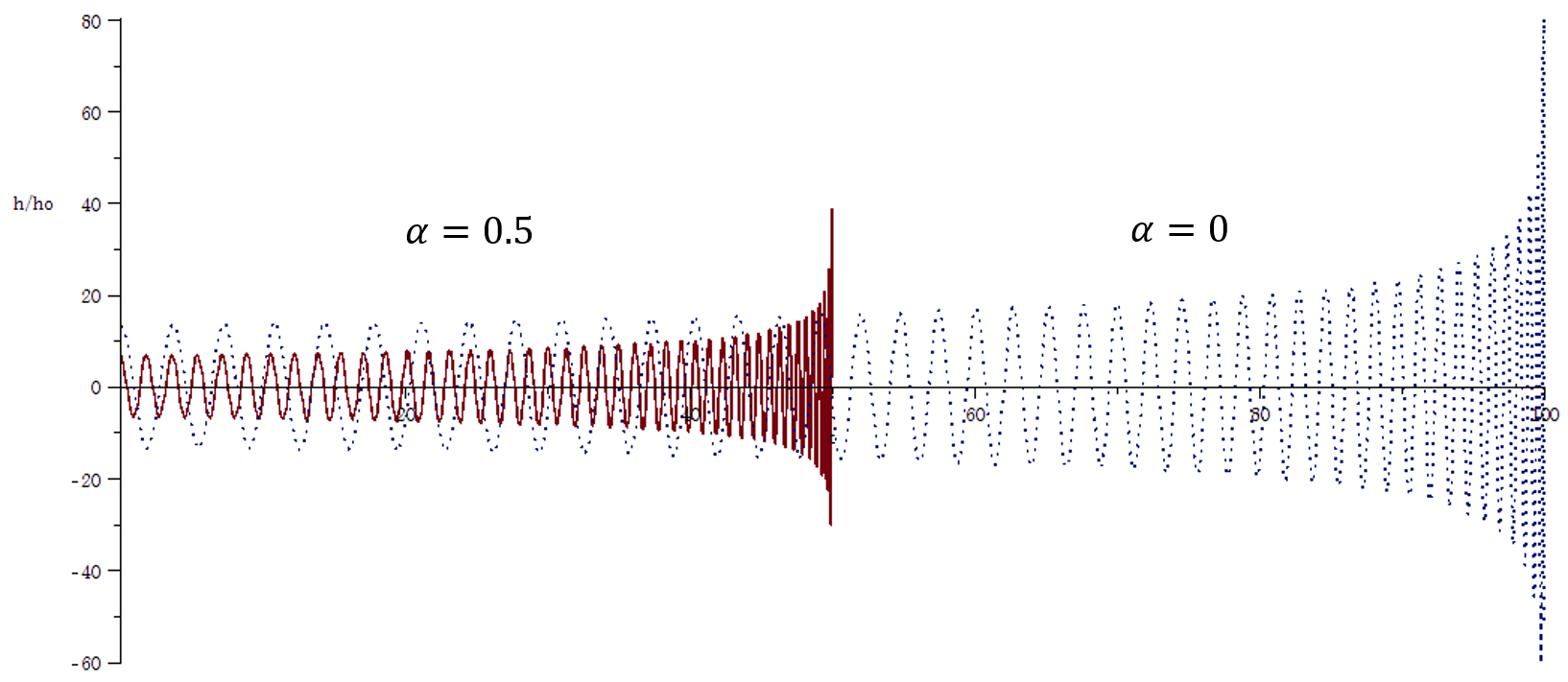}
\caption{Normalized Wave forms except the mass for wormhole binary system (solid) when $\alpha=0.5$ and normal star binary system (dotted) when $\alpha=0$}
\end{figure}

\begin{center}
\begin{figure}
{{\includegraphics[width=1.63in, height=1.81in, keepaspectratio=false]{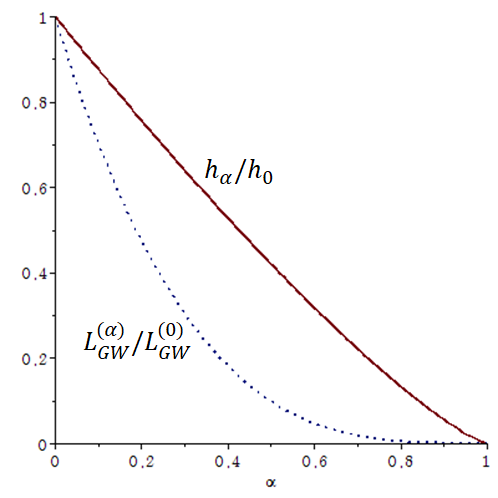}}}
{{\includegraphics[width=1.63in, height=1.81in, keepaspectratio=false]{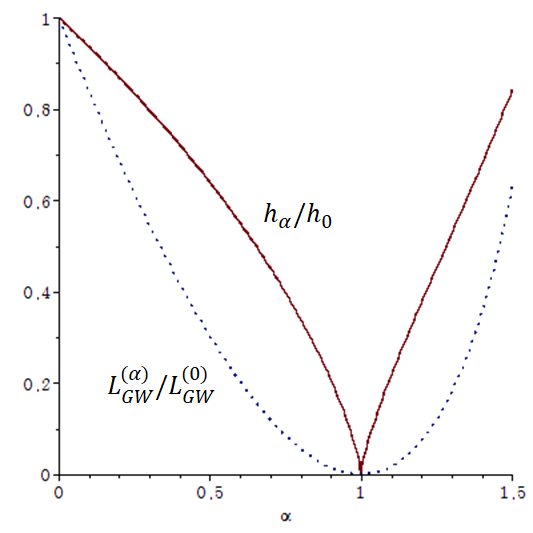}}}
\caption{Ratios of wave amplitude (solid line) and luminosity (dotted line) of twin wormhole-binary (left panel) and wormhole-black hole binary (right panel)
with respect to $\alpha$. All are normalized to the cases of twin black hole-binaries. Here, $\alpha=0$ means a black hole and $\alpha>1$ means a wormhole with negative effective mass.}
\end{figure}
\end{center}

\subsubsection{Wormhole-black hole binary}

Among the problems about wormhole-related binary systems with an effective mass definition, heterogeneous binary system such as wormhole-black hole binary is also interesting to us.
Since the detailed precess and future of the coalescence of heterogeneous binary system are unknown, here we will only deal with
the generation of gravitational waves due to orbital motion and the inspiral state before coalescence.

If only the wormhole and the black hole are considered pointlike particles and are sufficiently far away from each other, the whole processes is similar to the case of ordinary matter binary system.
As long as the effective mass of the wormhole is constant, the problem is similar to twin wormhole system when two wormholes approach each other.
In this case the mass and the gravitational potential are sufficient  as the physical quantities for gravitational wave generation.
Effective mass is needed for wormholes and not for black holes.

Let the mass of a black hole be $m_1=m$, the wormhole mass be $m$, and its effective mass be $m_2=m(1-\alpha)$.
When $\alpha <1$, the effective mass of the wormhole is also positive, so the binary system has the bounded orbital motion.
If $1<\alpha<2$, then the effective mass of wormhole is negative, but the total mass of this heterogeneous binary system is positive. The reduced mass is negative. As mentioned above, if the total energy is between zero and the maximum value of the effective potential, the motion is bounded.
If $2<\alpha$, the total mass is negative and the reduced mass is positive, then the motion of the binary system is not bounded.
The mass-related quantities are
\[
M_{wb}=m(2-\alpha), \quad \mu_{wb}=\f{m(1-\alpha)}{(2-\alpha)}, \quad
\eta_{wb}=\f{(1-\alpha)}{(2-\alpha)^2}.
\]

When $m_2$ is negative, $\mu$ and $\eta$ are also negative. Because the gravitational energy loss is given as positive-definite through $\eta^2$ and the total energy is positive, the relation $L_{GW} = -dE/dt$ stands for the negative energy radiation from this system.
The velocity change rate (\ref{eq:velocity}) is positive definite
due to the attractive interaction,
and so the $\eta$ in this relation should be replaced by $|\eta|$ for this system.

Thus the chirped mass is given by
\[
{\cal M}_{wb}=\left[ \f{|1-\alpha|^3}{(2-\alpha)} \right]^{1/5}m.
\]
Fig. 4 shows the normalized waveforms for wormhole-black hole binary system with waveform for black hole binary system assuming that the black hole mass is equal to the wormhole mass. For $\alpha=1.5$, i.e., a negative effective mass wormhole, the coalescence time is shorter but the amplitude is larger. For $\alpha=0.5$, i.e., a positive effective mass wormhole, the waveforms are similar to the wormhole binary system except that the coalescence time is longer. For negative effective mass, the amplitude and luminosity increase as $\alpha$ increases.
The ratio of the luminosity of this black hole-wormhole binary to the luminosity of the twin black hole binary is
\be
\f{L_{GW}^{(\alpha)}}{L_{GW}}= \f{(1-\alpha)^2}{(1-\alpha/2)^{2/3}}.
\ee
If the binary masses are different, there is a mass difference effect on this ratio.
The right panel of Fig. 3 shows the ratio of wave amplitude and luminosity of a wormhole-black hole binary to those of twin black hole binary as a function of $\alpha$.
As $\alpha1$ increases, the amplitude and luminosity decrease. However, for $1< \alpha <2$, the amplitude and luminosity increase as $\alpha$ increases. This is also shown in the waveform in Fig. 4.


\begin{figure}
\includegraphics*[width=8.29cm, height=4.79cm,
keepaspectratio=false]{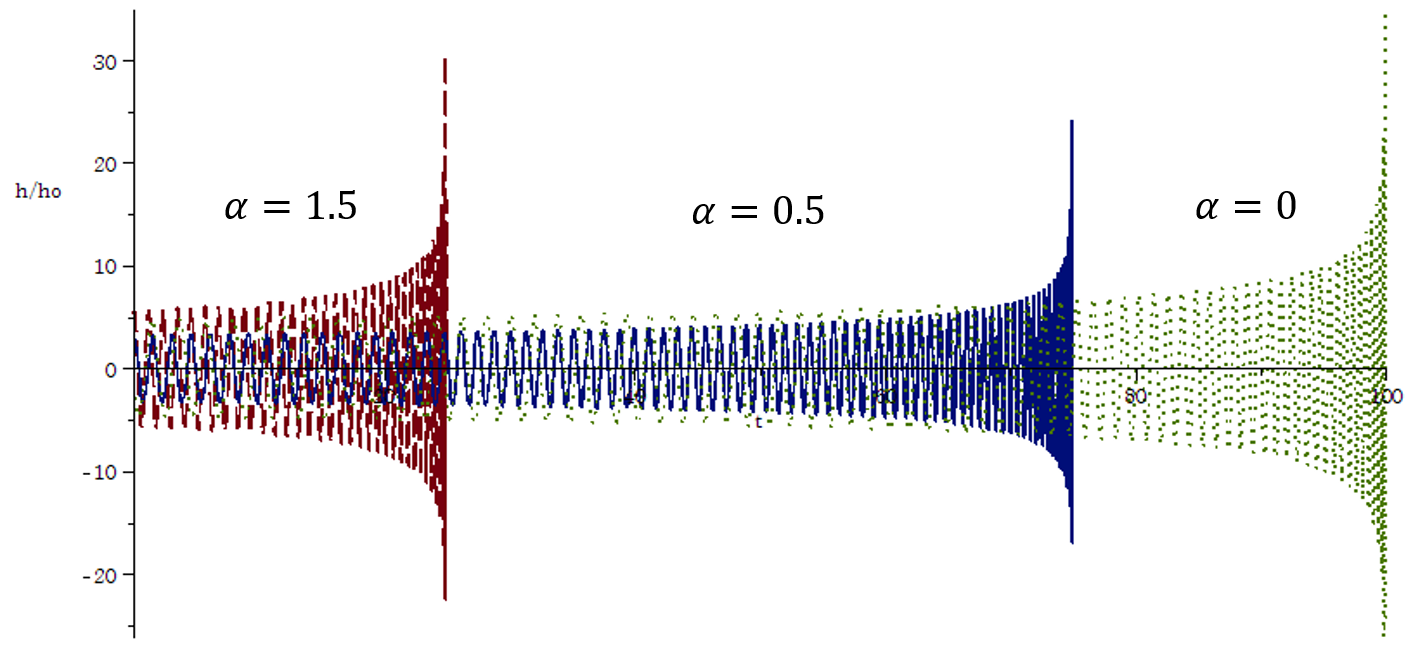}
\caption{Normalized waveforms for wormhole-black hole binary. Solid line is for $\alpha=0.5$, dashed line is for $\alpha=1.5$, and dotted line is for black hole binary ($\alpha=0$).}
\end{figure}

\subsubsection{Particles moving near a wormhole}

We will consider two cases: gravitational waves emitted by  a particle orbiting around a wormhole, and those by a particle moving linearly toward the center of the wormhole. In both cases, we limit the effective mass of the wormhole to a positive value and can set as $m(1-\alpha)$ with $\alpha<1$.
For the first case, we see the gravitational radiation by a particle of mass $m_0$ orbiting a wormhole of effective mass $m_{\mathrm{eff}}$.
When the particle orbits a wormhole, we can treat the binary system as a limit case where the mass of a particle is very small.
When $m_1=m_0 \ll m_2 = m_{\mathrm{eff}}$, the other mass-related quantities are
\[
\eta \rightarrow 0, \quad M \rightarrow m_{\mathrm{eff}}, \quad
\mu \rightarrow m_0.
\]
In this extreme case, the particle starts out with the velocity of $v_0$ and rotates around the wormhole, and the wormhole does not move. The particle then generate the gravitational waves and thereby loses energy.
As the particle's orbital radius gradually decreases and approaches the wormhole throat, $r_c$, the velocity also gradually increases, increasing to $v_c$. The time from the initial position to the throat is
 \[
 t_c = \f{5}{4096}\f{c^5}{G^3}\f{1}{m_0 m_{\mathrm{eff}}^2}(r_0^4-r_c^4).
 \]
 The energy loss rate is
 \[
 \int_0^{t_c}\f{dE}{dt}dt = \f{1}{2}m_0(v_c^2-v_0^2)=Gm_0m_{\mathrm{eff}}\left( \f{1}{r_c}-\f{1}{r_0} \right),
 \]
 where $v=\sqrt{2Gm_{\mathrm{eff}}/r}$ is the tangent velocity of the small particle.
The remaining energy is not enough to go further, so it cannot escape from the wormhole throat and eventually stops, losing energy.

As the second example, consider a case where a particle travels straight into a wormhole and passes through the throat. We start from $E=0$. The particle releases energy through the generation of gravitational waves and has negative energy, resulting in the bounded motion.
Assume that the particle's first path is from $\infty$ to $b$ and from $b$ through the throat to $R$ on the other side.

Analogous to the black hole case \cite{MM}, we consider a particle of mass $m_0$ starting from infinity in the positive direction of the $z$ axis with zero velocity,
\[
\f{1}{2}m_0 {\dot{z}}^2 - \f{Gm_0m_{\mathrm{eff}}}{z} = 0
\]
and the velocity is
\be
\dot{z}=-c \left( \f{R_s}{z} \right)^2,
\ee
where $R_s=2Gm_{\mathrm{eff}}/c^2$ is the Schwarzschild radius of the wormhole.
The rate of energy radiated through the gravitational wave is \cite{MM}
\be
L_{GW}=\f{2}{15}\f{Gm_0^2}{c^5}\left\langle(6\dot{z}\ddot{z}+2z\dddot{z})^2\right\rangle
\ee
Here the $z$ is the vertical axis through the wormhole center and the only non-vanishing component of the inertia tensor is
$I_{33}=m_0z^2$.
The particle in region I moves with $E=0$ from infinity to $b$ and to $R_1$ in region II of other side. It returned to mouth and to $R_2$ of the same side (region I) as the starting position, and continues these damped oscillatory motion until $R_n<b$ while $R_{n-1}>b$. If the $n$ is odd (even) number, $R_n$ is opposite (same) side. (Fig.~5).

\begin{figure}
\includegraphics*[width=8.29cm, height=5.2cm, keepaspectratio=false]{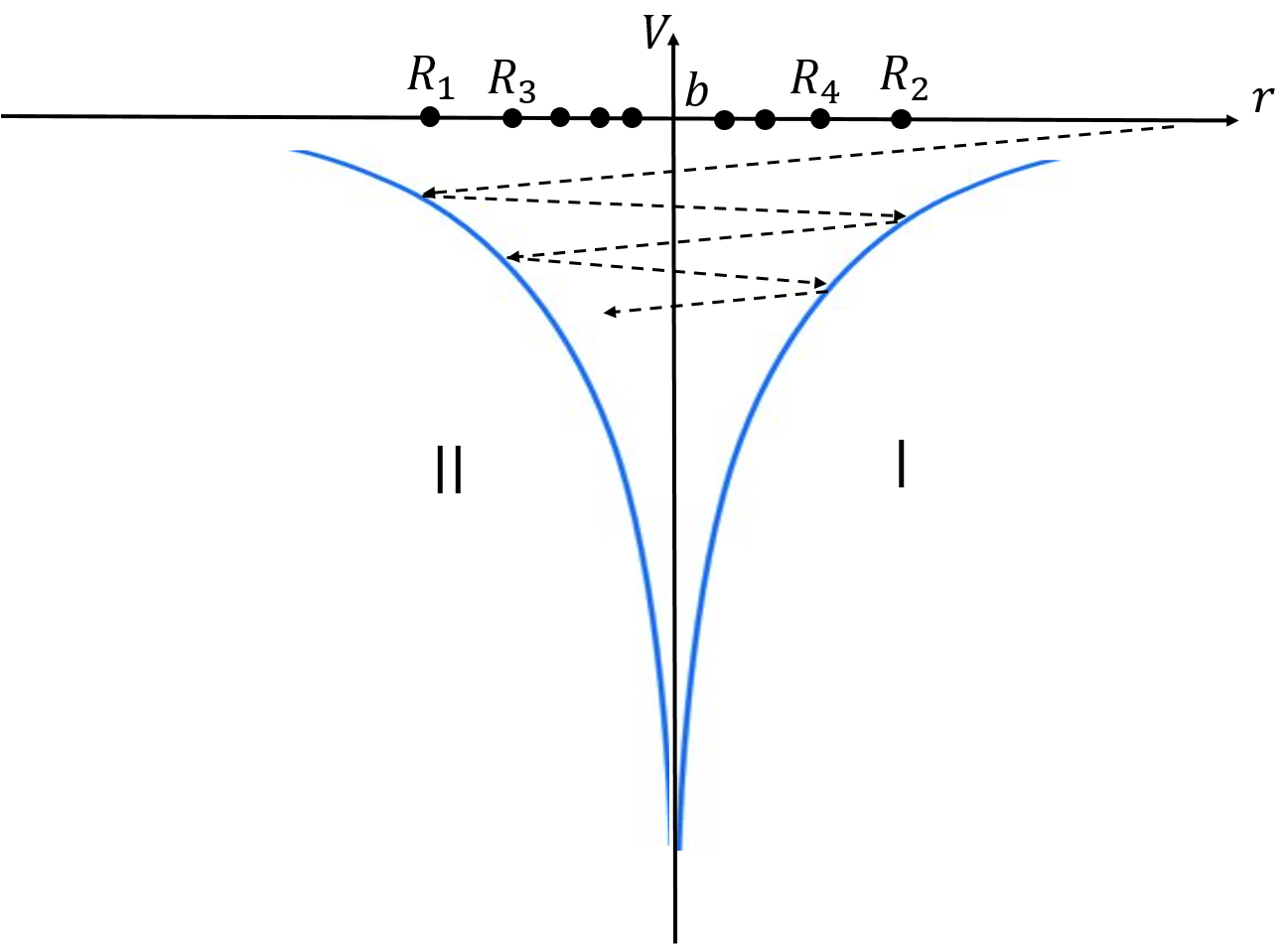}
\caption{The gravitational potential of a wormhole. The particle moves from infinity in region I with energy zero straight into the center and to $R_1$ in region II, $R_2$ in region I, $R_3$ in region II, etc. The particle passes through the throat ($b$) with each round trip.}
\end{figure}

The radiated energy falling into the wormhole from infinity to $b$ is
\[
E_0=\f{4}{105}\f{Gm_0^2}{R_s}\left( \f{R_s}{b} \right)^{7/2}
\]
and the energy radiated when moving from $b$ to $R_n$ is
\Be
E_n &=& \f{4}{105}\f{Gm_0^2}{R_s}\left[
\left( \f{R_s}{b} \right)^{7/2}-\left(\f{R_s}{R_n} \right)^{7/2} \right] \nonumber \\
&=& E_0\left( 1 - \gamma_n^{7/2} \right),
\Ee
where $\gamma_n = b/R_n$ and $0<\gamma_n<1$.
Thus the energy radiated by the particle travel from infinity to $b$ and $R_1$ in other side through throat is $E_0+E_1$. When it is repeated to $R_n$, the total energy radiated is
\be
E_{\mathrm{rad}}^{(n)}=E_0+2\sum_{k=1}^{n-1}E_k + E_n. \label{eq:energy}
\ee
Therefore, the particle sequentially radiates energy $E_n$ in alternating regions, so that
it release energy discontinuously rather than continuously.
As the oscillatory motion continues, the duration until $R_n$ on one side appears becomes shorter, and the radiant energy $2E_n$ emitted at that time becomes smaller than before.
We also can derive the position of $R_n$ by equalizing the potential energy
\be
\f{Gm_0m_{\mathrm{eff}}}{R_n}=E_0+2\sum_{k=1}^{n-1}E_k + E_n. \label{eq:pot_to_energy}
\ee
Using $\beta \equiv Gm_0m_{\mathrm{eff}}/(bE_0)$, which is the ratio of the potential energy at $b$ to the energy loss moving from infinity to $b$, $R_n$ can be obtained by solving the equation:
\be
\beta \gamma_n + \gamma_n^{7/2}=2n- 2 \sum_{k=1}^{n-1}\gamma_k^{7/2} \label{eq:gamma}
\ee
on the premise that we already get all $R_k, k<n$. Comparing the values of each term, $\gamma$'s are very small compare to $\beta$, which is nearly to the ratio of the wormhole mass and the test mass. The approximate solution for $R_n$ is
\be
R_n \simeq \f{\beta}{2n} b,
\label{eq:rad_distance}
\ee
by neglecting the $\gamma^{7/2}$ terms and the total energy loss is
\Be
E_{\mathrm{rad}}^{(n)} &=& \f{Gm_0m_{\mathrm{eff}}}{R_n} \simeq 2nE_0 \nonumber \\
&\simeq& \f{4}{105}n m_0c^2 \left( \f{m_0}{m_{\mathrm{eff}}} \right)
\Ee
by (\ref{eq:energy}) and (\ref{eq:pot_to_energy})
when $n$ is not so large and $b \simeq R_s$.
This is because for large $n$ the $(b/R_k)^{7/2}$ terms cannot be neglected, so they are added when estimating the energy loss.
When the number $n=\beta/2$, the particle stops at throat with the total energy loss is
the half of test particle's mass energy as
\be
E_{\mathrm{rad}} = \f{Gm_0m_{\mathrm{eff}}}{b} \simeq \f{1}{2}m_0 c^2. \label{eq:stop_energy}
\ee
Half of the energy (\ref{eq:stop_energy}) is radiated to region I, and the remaining energy is radiated to region II in discontinuous forms.
When $n$ becomes large, it makes two contributions to the equation: the property that $\gamma_n$ approaches 1 and the sum of $\gamma_k$.
If we define the sum as $s(n) \equiv \sum_{k=1}^{n}k^{7/2}$, the $n$-th distance becomes
\be
\gamma_n \simeq \f{2}{\beta}(n - \Delta n)
\ee
and the energy loss is
\be
E_{\mathrm{rad}}^{(n)} \simeq \f{4}{105} m_0c^2 \left( \f{m_0}{m_{\mathrm{eff}}} \right)
(n - \Delta n),
\ee
where
\[
\Delta n = \left( \f{2}{\beta} \right)^{7/2}\left(s(n)-\f{1}{2}n^{7/2} \right).
\]
Let $\alpha = 200$, $n=100$. Here the contribution of $s(n)$ is $2.27 \times 10^8$ and $\Delta n \simeq 22$. Therefore the number becomes 78 and the total energy loss is reduced by 22\%. Stopping at the throat requires more travel, but the total energy loss is equal to  (\ref{eq:stop_energy}),  half the mass-energy of test particle.

For the case of black hole, the particle is absorbed into the black hole with the radiation of \cite{MM}
\[
E \simeq 0.01m_0c^2\left( \f{m_0}{M} \right),
\]
where $M$ is the black hole mass.
The particle under the gravitation of a wormhole has damped oscillatory motion, emitting the discrete form of radiation until it is captured by the wormhole.

\section{Conclusion}

We considered a binary system containing wormhole(s) as a gravitational wave source. Applying the Newtonian approximation to this binary system, the wormhole effective mass definition is needed from the Newtonian potential of a wormhole.
Therefore when applying the interaction of any object with a wormhole, the gravitational potential of the wormhole should be used as the effective mass, including the tension term in Newtonian approximation. Here we tried to find the effective mass-related properties of gravitational wave generated by the binary system.
The invariance of the effective mass shows that the wormhole can be considered a point mass. If not, size-effects can be considered using  methods such as the effective one-body theory.

We can also see that the motion of the system depends on the sign of effective mass. Bounded motion is possible as long as the total mass is positive, even if one of the masses is negative.
We found the waveforms and energy losses for three cases: wormhole-wormhole, wormhole-black hole, and particle-wormhole. The effective mass changes the coalescence time and the waveforms of the binary system.

In the last two examples, the particles' motions are also different from that near a black hole where particles are absorbed. In the case of a wormhole, particles with excess energy can move to other region through the throat. Or the particles moving straight into wormhole will go into damped oscillatory motion between two regions, losing half of its mass energy, until it stops. In this case, different from the case of the particle absorbed by the black hole, the energy is radiated in a discrete and increasingly smaller form.

\acknowledgments

This work was supported by National Research Foundation of Korea (NRF) funded by the Ministry of
Education (2021R1I1A1A01056433).

\end{document}